\documentclass{emulateapj}
\usepackage[below]{placeins}
\usepackage{natbib}
\slugcomment{To be published by Astrophysical Journal Letters}

\def\gtorder{\mathrel{\raise.3ex\hbox{$>$}\mkern-14mu
    \lower0.6ex\hbox{$\sim$}}}
\def\ltorder{\mathrel{\raise.3ex\hbox{$<$}\mkern-14mu
    \lower0.6ex\hbox{$\sim$}}}

\def\kmsmpc{\ {\rm km~s^{-1} Mpc^{-1}}}
\def\msun{M_\odot}

\shorttitle{Bar Triggering by Interactions with DM}
\shortauthors{}

\begin{document}

\title{Disk Evolution and Bar Triggering Driven by Interactions with 
Dark Matter Substructure
}

\author{ 
Emilio Romano-D\'{\i}az\altaffilmark{1},
Isaac Shlosman\altaffilmark{2,1},
Clayton Heller\altaffilmark{3},
Yehuda Hoffman\altaffilmark{4}
}
\altaffiltext{1}{
Department of Physics and Astronomy, 
University of Kentucky, 
Lexington, KY 40506-0055, 
USA
}
\altaffiltext{2}{
JILA, 
University of Colorado, 
Boulder, CO 80309, 
USA
}
\altaffiltext{3}{
Department of Physics, 
Georgia Southern University, 
Statesboro, GA 30460, 
USA
}
\altaffiltext{4}{
Racah Institute of Physics, Hebrew University; Jerusalem 91904, Israel
}

\begin{abstract}
We study formation and evolution of bar-disk systems in fully self-consistent
cosmological simulations of galaxy formation in the $\Lambda$CDM WMAP3
Universe. In a representative model we find that
the first generation of bars form in response to the asymmetric dark matter
(DM) distribution (i.e., DM filament) and quickly decay. Subsequent bar
generations form and are destroyed during the major merger epoch permeated by 
interactions with a DM substructure (subhalos). A long-lived bar is triggered
by a tide from a subhalo and survives for $\sim 10$~Gyr. The evolution of this
bar is followed during the subsequent numerous minor mergers and interactions
with the substructure. Together with intrinsic factors, these interactions
largely determine the stellar bar evolution. The bar strength and its 
pattern speed anticorrelate, except during interactions and when the
secondary (nuclear) bar is present. For about 5~Gyr bar 
pattern speed {\it increases substantially} despite 
the loss of angular momentum to stars and {\it cuspy} DM halo. We analyze the
evolution of stellar 
populations in the bar-disk and relate them to the underlying dynamics. While 
the bar is made mainly of an intermediate age, $\sim 5-6$~Gyr, disk stars at 
$z=0$, a secondary nuclear bar which surfaces at $z\sim 0.1$ is made of younger, 
$\sim 1-3$~Gyr stars. 
\end{abstract}

\keywords{cosmology: dark matter --- galaxies: evolution --- galaxies:
formation --- galaxies: halos --- galaxies: interactions --- galaxies:
kinematics and dynamics}
    
\section{Introduction and Numerics}
\label{sec:intro}

Within the framework of structure formation in the universe, a hierarchy
of dark matter (DM) masses form, while baryons assemble in their midst
as galactic disks (e.g., White \& Rees 1978). Disk evolution is expected
to be influenced heavily by mergers and interactions in various
ways. In this Letter, we focus on the specific issue of a tidal triggering 
of stellar bars embedded in a disk and surrounded by a DM halo with a
substructure, within the context of cosmological evolution in the WMAP3 universe
(see also Dubinski et al. 2008). We analyze the dynamical aspects of this 
evolution and follow the prevailing stellar populations in the bar-disk system.    

Recent efforts to understand galaxy formation have been spearheaded by the 
work on pure DM halo formation (e.g., Diemand et al. 2007) and 
associated disk growth (e.g., Sommer-Larson et al. 2003; Governato et al. 
2004, 2007; Heller et al. 2007a,b). Addition of baryonic component(s) to this 
modeling has met with difficulties, partly due to a numerical resolution, star 
formation (SF), energy feedback, and other processes.
Dissipation is relatively slow in disks, but can be accelerated
by any substantial departure from axial symmetry due to gravitational 
torques --- a nonlocal viscosity, surpassing the conventional viscosity by 
orders of magnitude. Torques facilitate the redistribution 
of mass, angular momentum and energy between the disk baryons and the slowly 
tumbling DM halo. Two main sources can serve as triggers of gravitational torques 
onto the disk --- an asymmetric mass distribution in the
DM halo, and its non-uniformity, i.e., substructure, in DM and baryons.  
Both tidal interactions (e.g., Byrd et al. 1986; Noguchi 1987; Gerin et al. 1990)
and DM halo asymmetry (Heller  et al. 2007a,b) are likely to trigger stellar 
bars. About 1/3 of the large-scale bars host additional nuclear bars (Laine 
et al. 2002; Erwin \& Sparke 2002).     

Observations and numerical simulations point to an intricate relation between
the stellar populations and the underlying dynamics of bars (e.g., Martin \& Friedli
1997; Knapen et al. 1995a,b; Jogee et al. 2002a,b; Zurita \& Perez 2008). For 
bars in a cosmological context, this issue becomes even more relevant, because 
the disk is open to accretion of cold gas and energy input from 
interactions. Here we attempt to answer some questions about the long-term 
evolution of bar-disk populations and relate it to the dynamical history of the
system.  

So far, nearly all modeling of bar evolution has been limited to isolated systems 
in a stationary state, with the exception of simulations of tidal interactions in 
controlled experiments (e.g., Gauthier et al. 2006). We test the bar 
origin and evolution in the cosmological setting without fine-tuning and minimazing
the prior assumptions.

Numerical simulations have been performed using the FTM-4.5 hybrid $N$-body/SPH 
code (e.g., Heller \& Shlosman 1994; Heller et al. 2007b; 
Romano-Diaz et al. 2008a) using physical coordinates. The total number of DM 
particles is $2.2\times 10^6$
and the SPH particles $4\times 10^5$. The gravity is computed using the falcON 
routine (Dehnen 2002) which scales as O(N). The gravitational softening is
500~pc, for DM, stars and gas. We assume the $\Lambda$CDM cosmology with
WMAP3 parameters, $\Omega_m=0.24$, $\Omega_\Lambda=0.76$
and $h=0.73$, where $h$ is the Hubble constant in units of $100~\kmsmpc$. The
variance $\sigma_8=0.76$ of the density field convolved with the top hat window
of radius $8h^{-1}$~Mpc is used to normalize the power spectrum.
The SF algorithm is described in Heller et al. (2007b). We note
only that multiple generations of stars can form from a single SPH particle. 
The energy and momentum feedback into the ISM are 
implemented, and the associated parameters have the following 
values (Heller et al.): energy thermalization $\epsilon_{\rm SF}=0.3$, cloud
gravitational collapse $\alpha_{\rm ff}=1$, and self-gravity fudge-factor 
$\alpha_{\rm crit}=0.5$. 

Initial conditions generated here are those of Romano-Diaz et al. (2008a): we use
the Constrained Realizations (CR) method (Hoffman \& Ribak 1991) within a restricted
box size of $8h^{-1}$~Mpc and a sphere of $5h^{-1}$~Mpc is carved out and evolved. 
A large-scale filament runs across the computational sphere and is banana-shaped.
The constructed Gaussian field is required to obey a set of constraints
of arbitrary amplitudes and positions (more about this method in Romano-Diaz
et al. 2006, 2007). Two constraints were imposed on the density field, first ---
that the linear field Gaussian smoothed with a kernel of 
$1.0\times 10^{12}~h^{-1}\msun$, has an over-density of $\delta=3$ at the origin
($2.5\sigma$ perturbation,  
where $\sigma^2$ is the variance of the appropriately smoothed field). It was
imposed on a $256^3$ grid and predicted to collapse at $z_{\rm c}\sim 1.33$
based on the top-hat model. This perturbation  
is embedded in a region (2nd constraint) corresponding to a mass of 
$5\times 10^{13}~h^{-1}\msun$ in which the over-density is zero, i.e.,
the unperturbed universe. The random component in CRs favors formation of similar
structures which leads to major mergers. The DM mass inside the computational sphere is
$\sim 6.1\times 10^{12}~h^{-1}\msun$. We have randomly
replaced 1/6 of DM particles by equal mass SPH particles. Therefore, 
$\Omega_{\rm m}$ is not affected.

\section{Evolution and Stellar Populations}

The prime halo evolves through a series of major mergers which end by $z\sim
1.5$, while accretion of subhalos and smooth accretion continue to the present, 
mainly along the filament. The number of subhalos varies with time, and at its 
peak is about few hundred, for subhalos in excess of $10^8~\msun$. The disk 
is recognizable from $z\sim 8$, associated
with a strong SF. It grows from inside out and 
appears strongly non-axisymmetric and barred along the filament. The first 
merger of nearly equal masses is gas-rich, coplanar and prograde. The disk
reforms immediately, although the bar does not ($z\sim 5$). Next merger, at 
$z\sim 4$, is again prograde and excites a bar which is destroyed shortly. 
Subsequent minor mergers drive
a strong spiral structure, as the disk grows to $\sim 10$~kpc, and push the gas 
inwards in a `shepherding' mode. The SF is vigorous in the high surface density 
gas. At $z\sim 2.6$, the tide from a subhalo (prograde and inclined 
penetrating minor merger) triggers a strong gas-rich bar (Fig.~1) followed by
(minor) merger activity till $z=0$. A direct hit by another disk (in a host 
subhalo), on $\sim 40^\circ$-inclined prograde trajectory, weakens
the bar abruptly at $z\sim 2.2$ and shortens it. A retrograde encounter at 
$z\sim 1.4$ has no dramatic effect on the bar. As a result of these 
interactions the gas-rich disk shrinks gradually. 
Overall, the disk axis is oriented along the halo minor axis, but this
orientation experiences strong departures during the major merger epoch.
The DM halo forms an isothermal cusp, $R^{-2}$, within 15~kpc which survives
beyond $z\sim 1$ and is ultimately washed out, leaving a flat core of 
$\sim 2-3$~kpc (Romano-Diaz et al. 2008b).

\begin{figure*}
\begin{center}
\includegraphics[angle=0,scale=0.25]{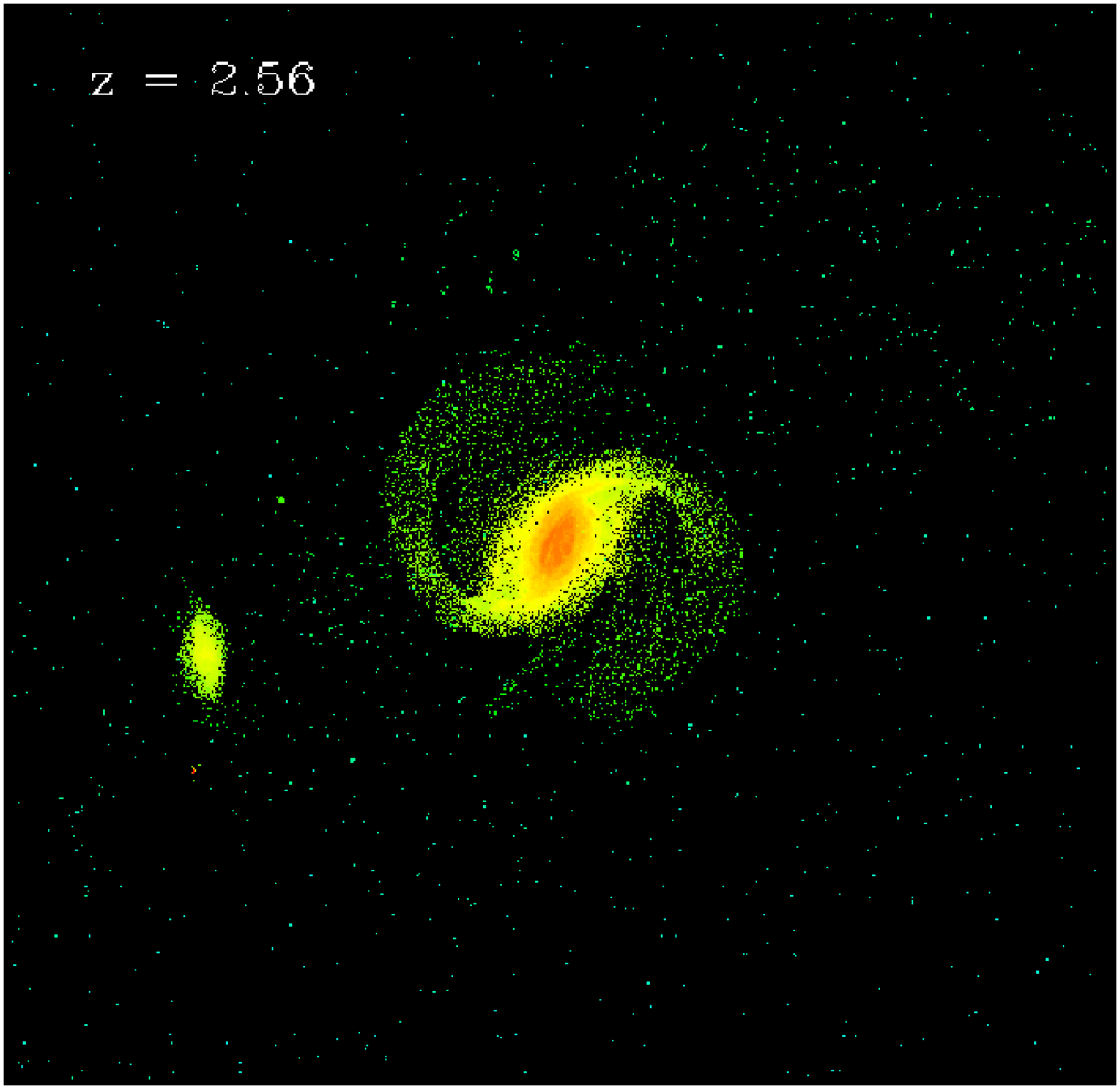}
\includegraphics[angle=0,scale=0.25]{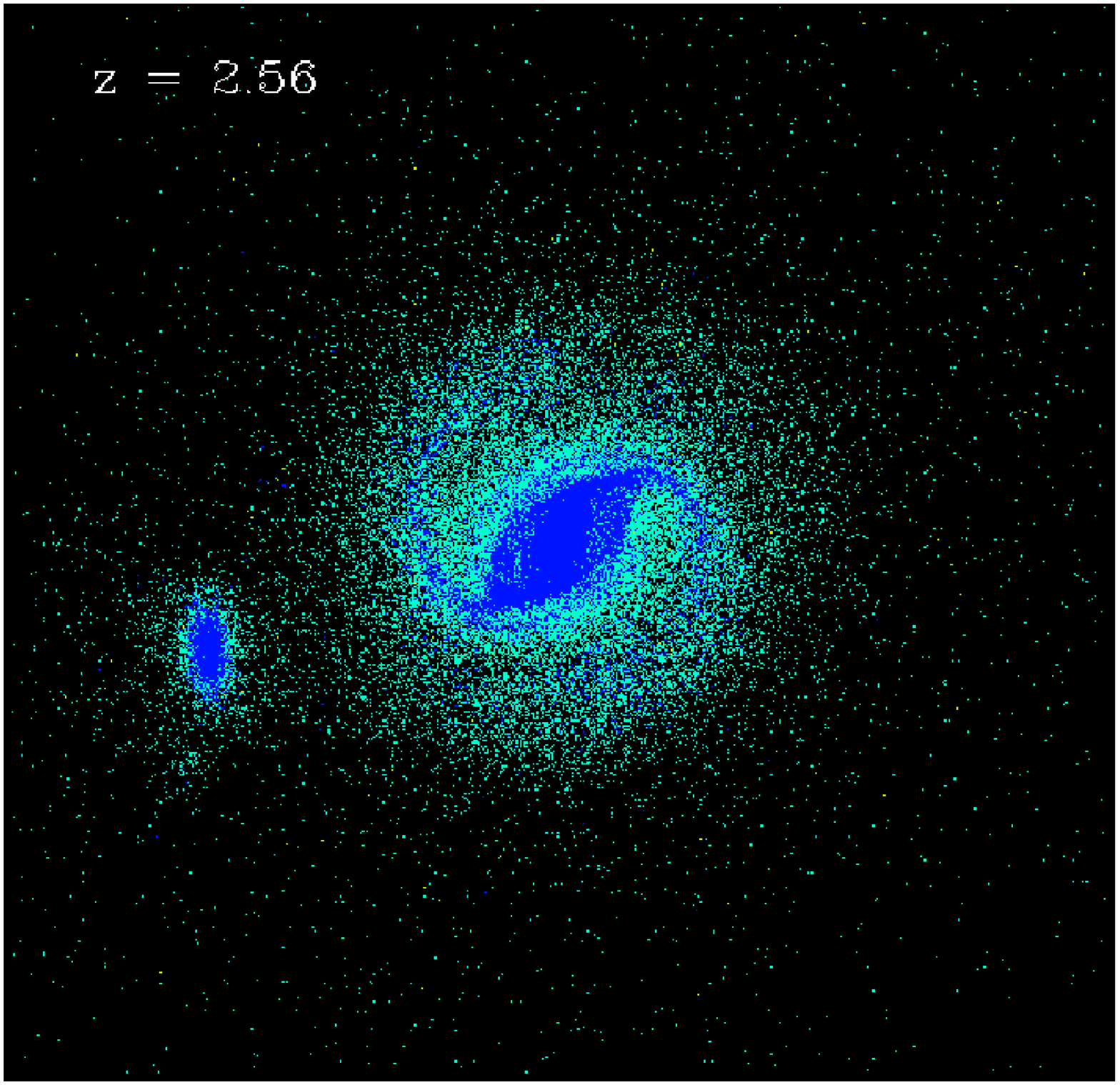}
\end{center}
\caption{Penetrating minor merger triggering gas-rich stellar bar in the
growing disk. The intruder is on the prograde trajectory inclined to the
disk plane by about $50^\circ$ with an impact parameter of $\sim 9$~kpc.
{\it Left:} gas; {\it Right:} stars --- the color palette traces the stellar
ages (0--0.2~Gyr blue; 0.2 -- 2~Gyr aquamarine). The intruder is also gas rich
and is shown near the closest first approach.
}
\end{figure*} 

Bar dynamics is displayed in Fig.~2. The pattern speed, $\Omega_{\rm b}$, of 
the initial barlike response to the DM stagnates initially, then accelerates.  
The bar amplitude, $A_2$ of the $m=2$ Fourier mode,
decays rapidly. The oval perturbation in the stellar disk remains but is
weak and noisy, and therefore omitted here until the tidal
triggering at $z\sim 2.6$. Subsequent evolution depends both on interactions 
with the substructure and on intrinsic factors.

\begin{figure}
\begin{center}
\includegraphics[angle=0,scale=0.4]{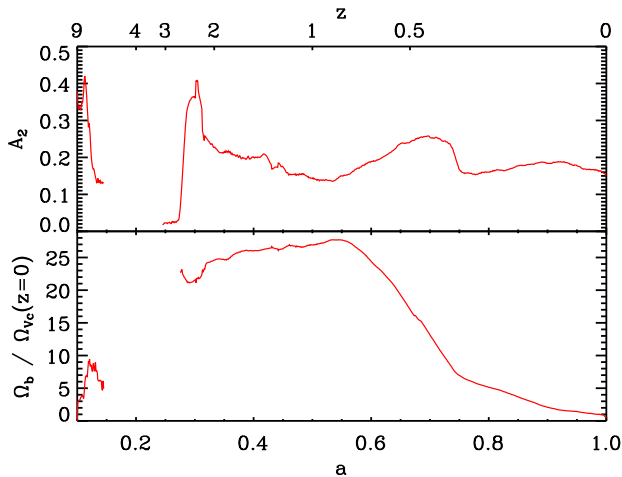}
\end{center}
\caption{Evolution of the bar amplitude $A_2$ (upper) and pattern speed 
$\Omega_{\rm b}$ (lower) with the cosmological expansion factor $a$ and redshift $z$. 
The gap corresponds to a particularly intensive merger activity when both curves are 
ill-defined. $\Omega_{\rm b}$ is normalized by the angular velocity of a test
particle at the radius of a maximal rotation velocity in the halo at $z=0$.
The first generation bar (during the first Gyr) is induced by the main 
DM filament, while the next generations are tidally-induced and destroyed. Finally,
a long-lived bar is tidally triggered by the interaction 
with a subhalo in a a highly-inclined prograde encounter (Fig.~1).
}
\end{figure} 

\begin{figure*}
\begin{center}
\includegraphics[angle=0,scale=0.14]{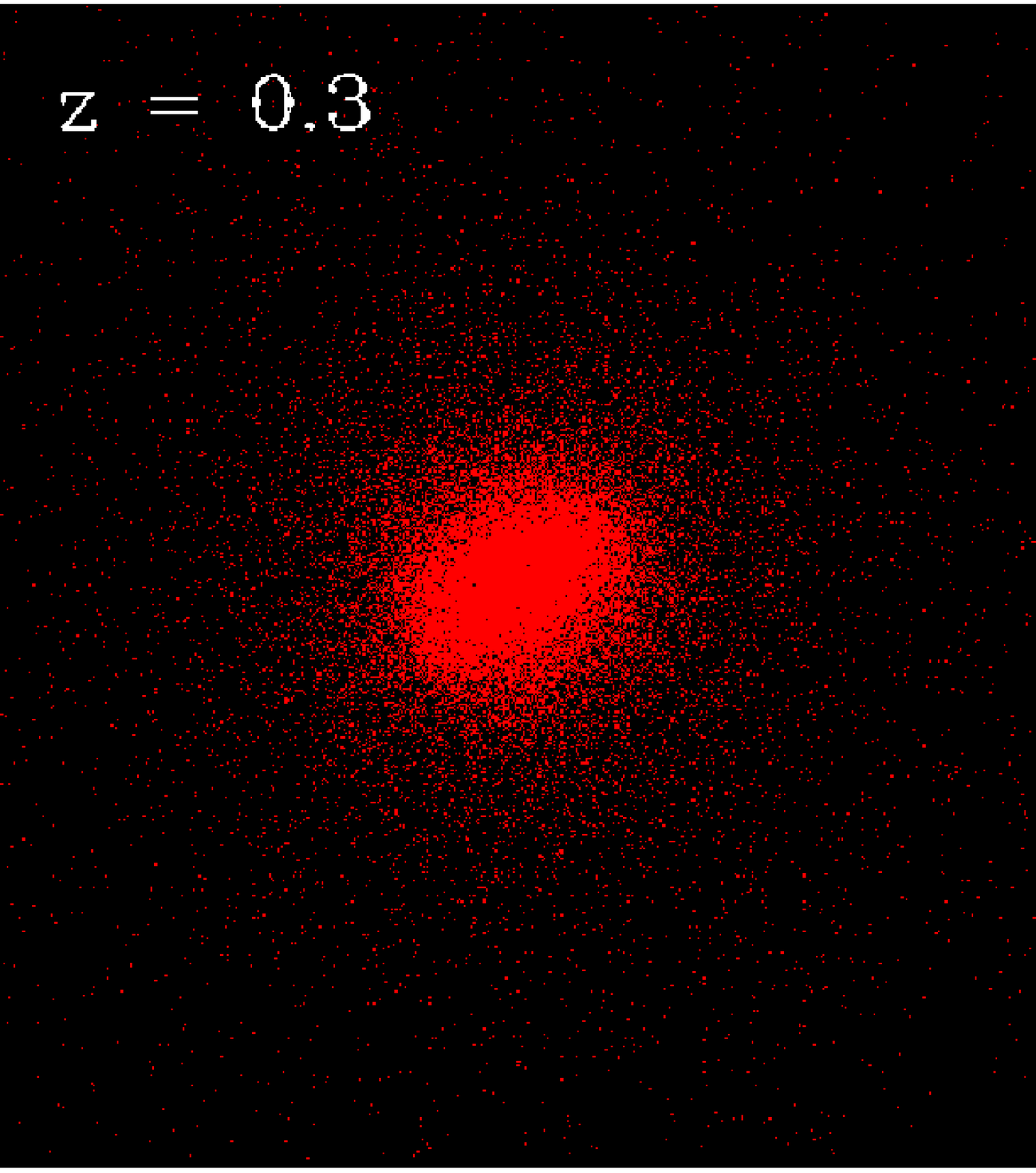}
\includegraphics[angle=0,scale=0.14]{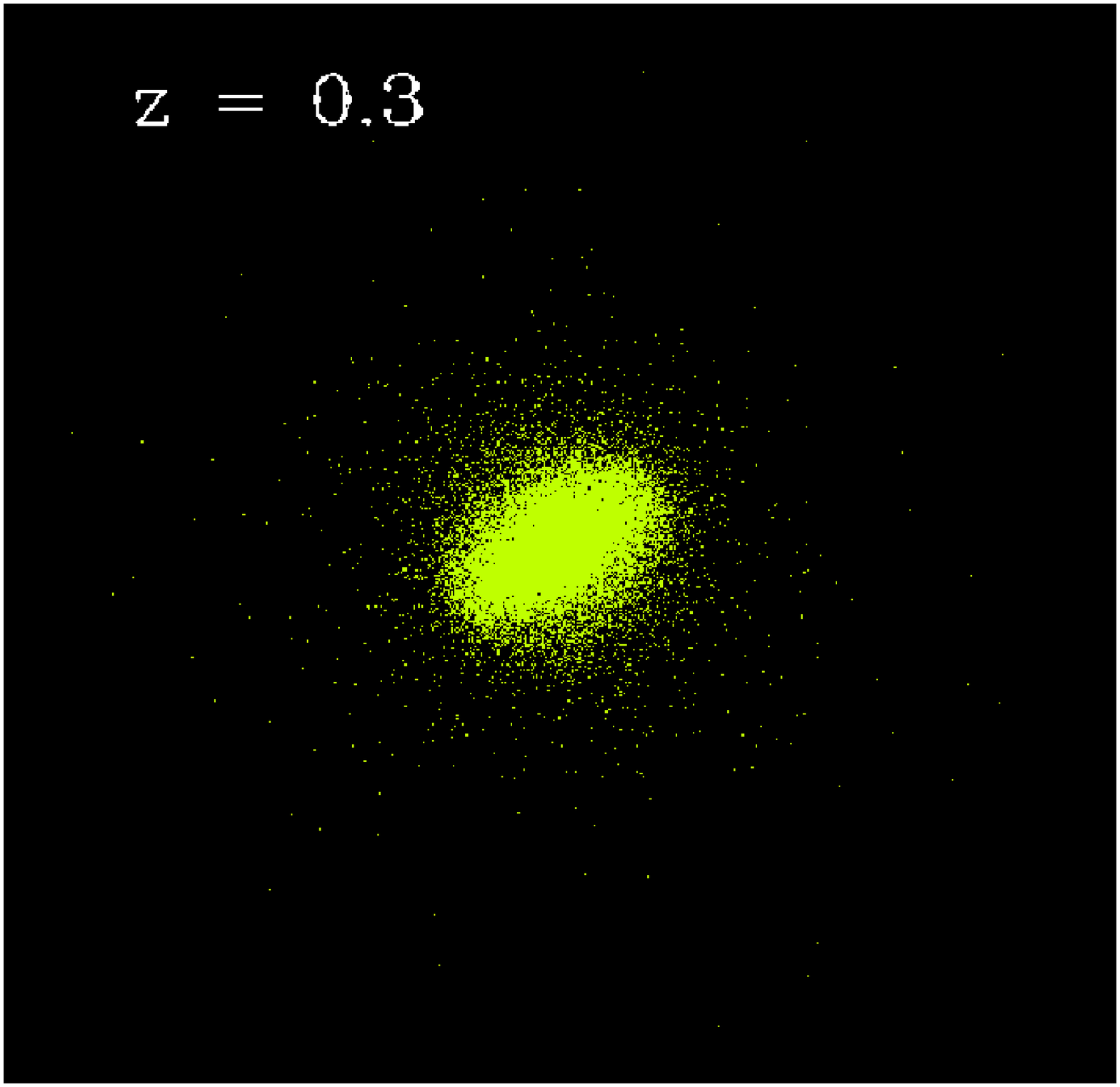}
\includegraphics[angle=0,scale=0.14]{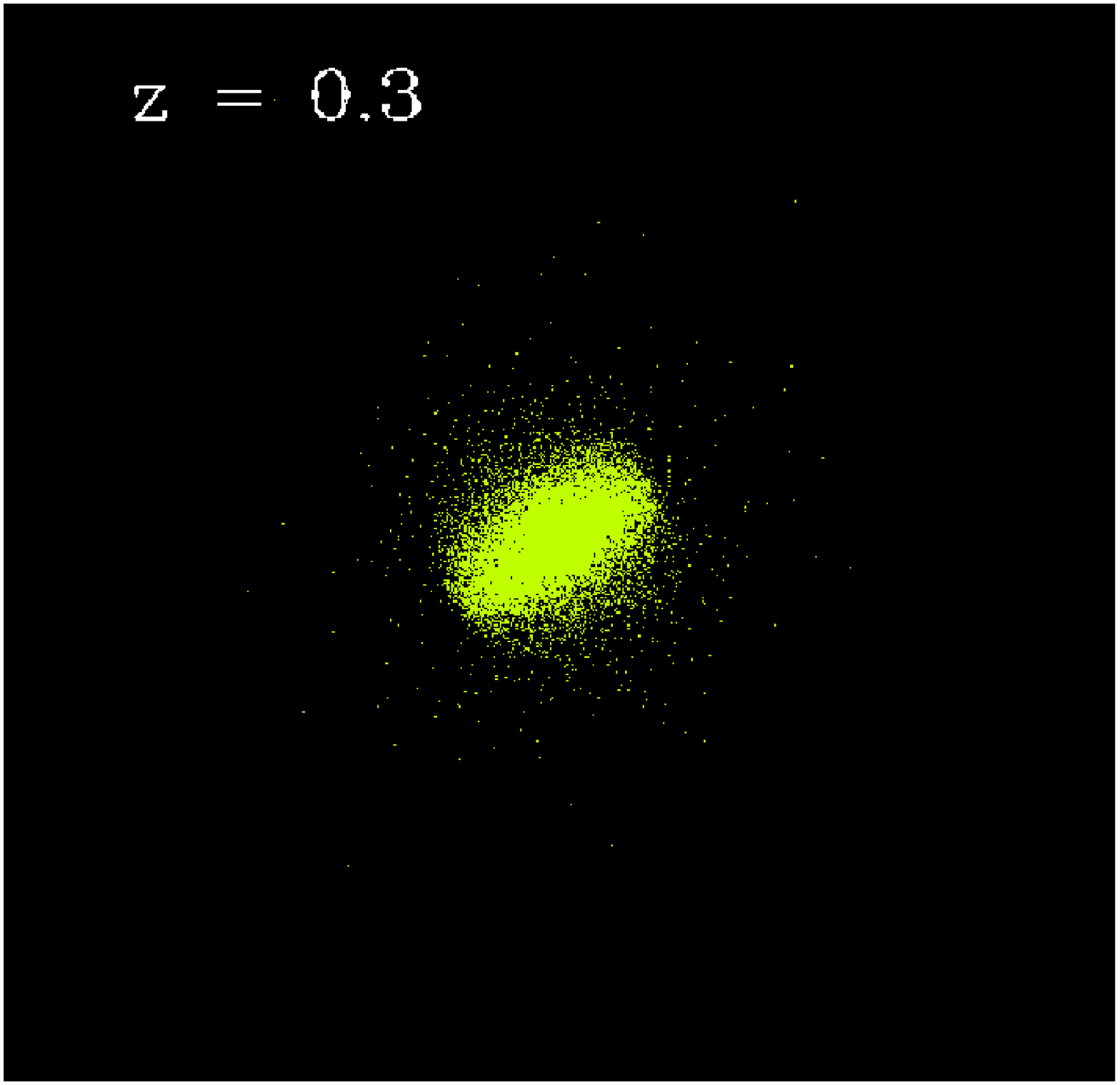}
\includegraphics[angle=0,scale=0.14]{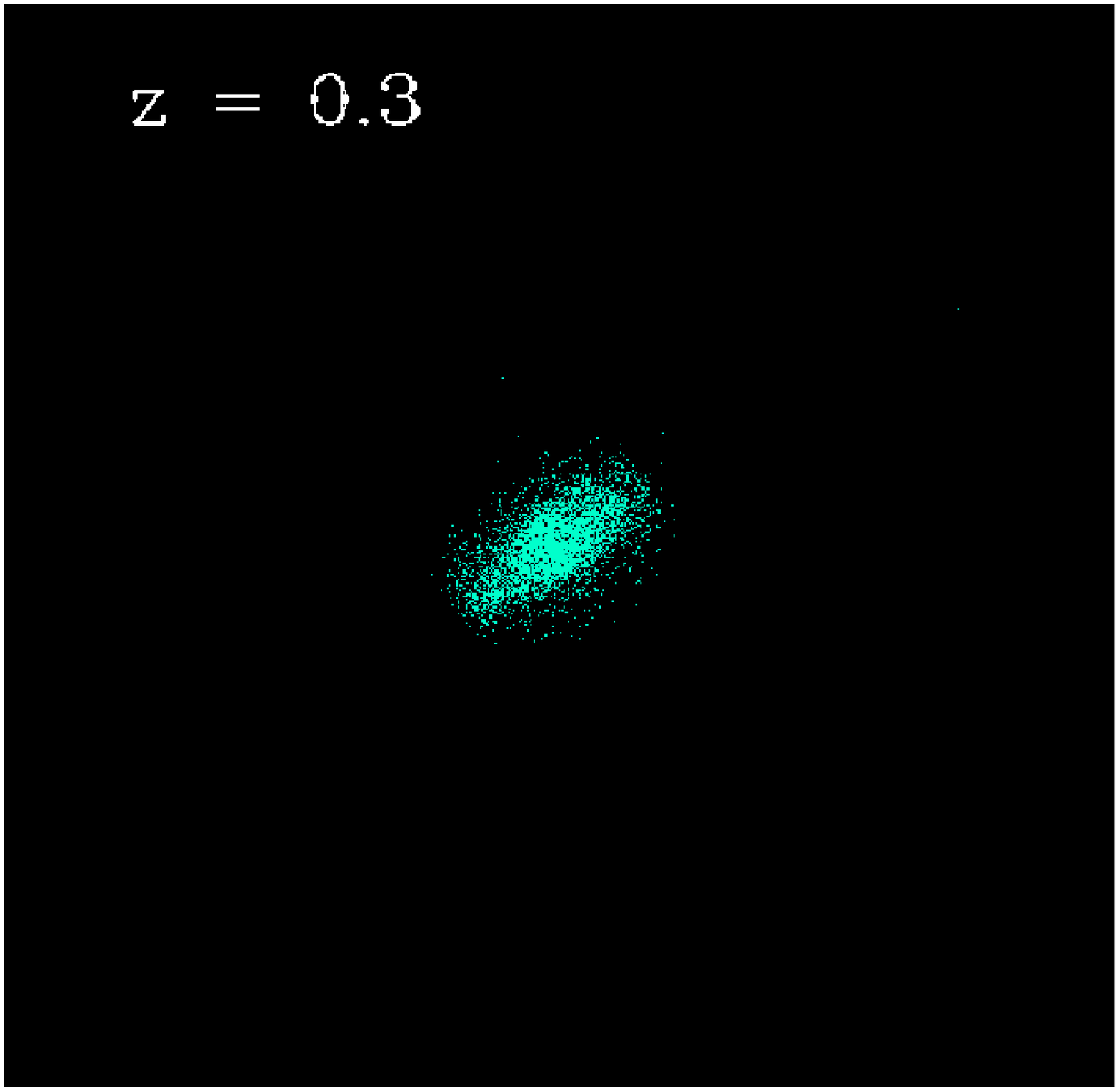}
\includegraphics[angle=0,scale=0.14]{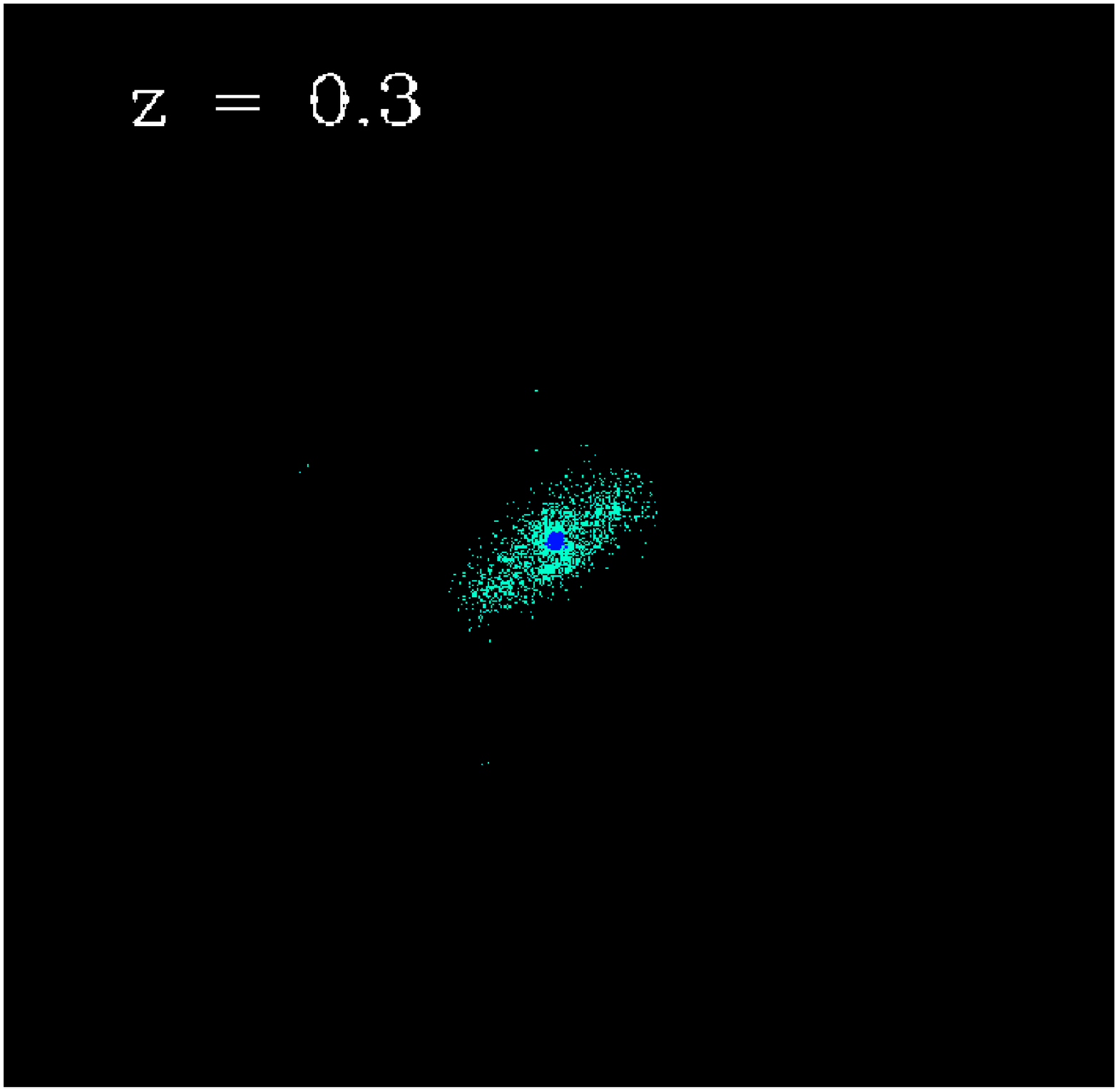}
\includegraphics[angle=0,scale=0.14]{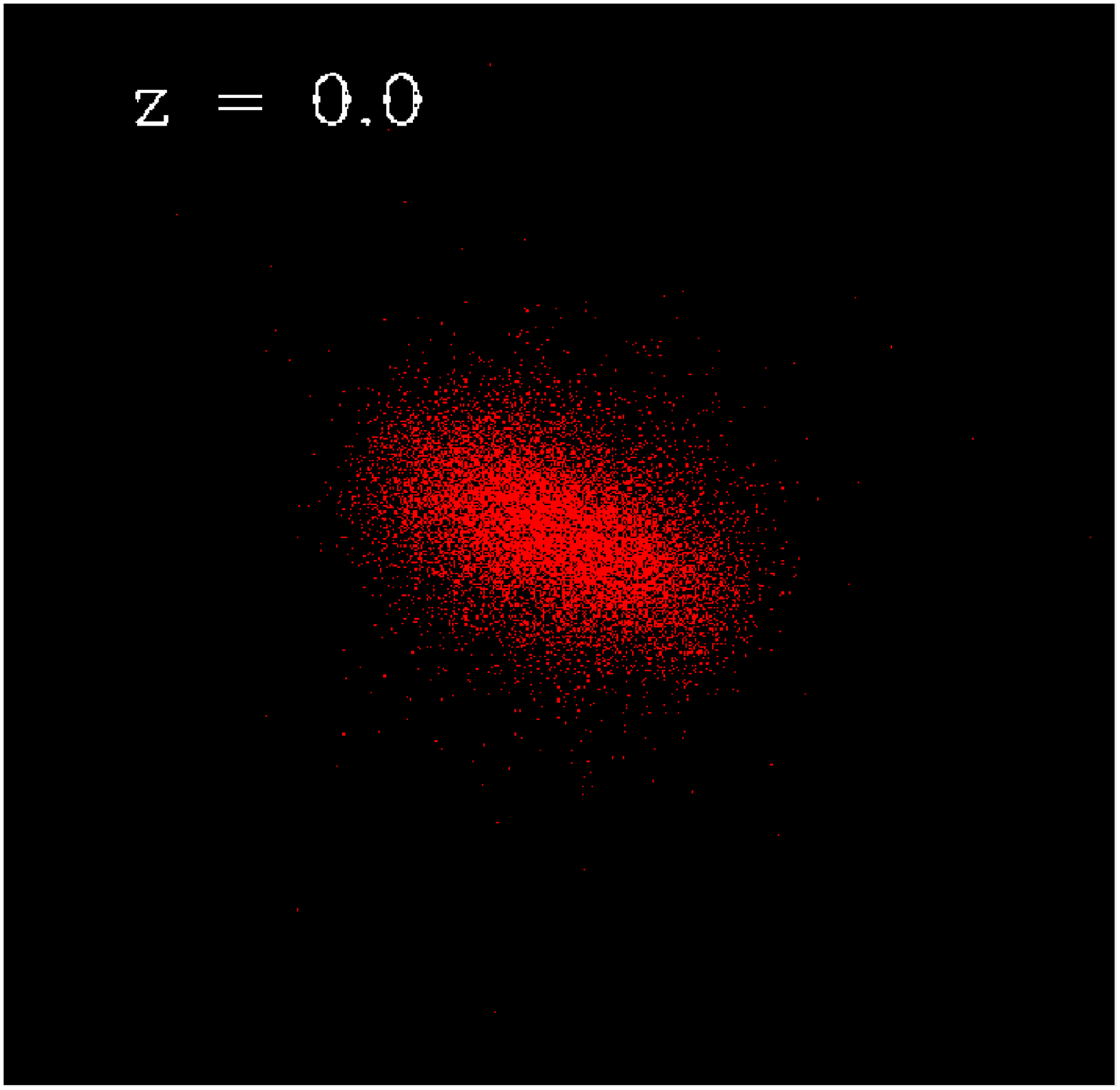}
\includegraphics[angle=0,scale=0.14]{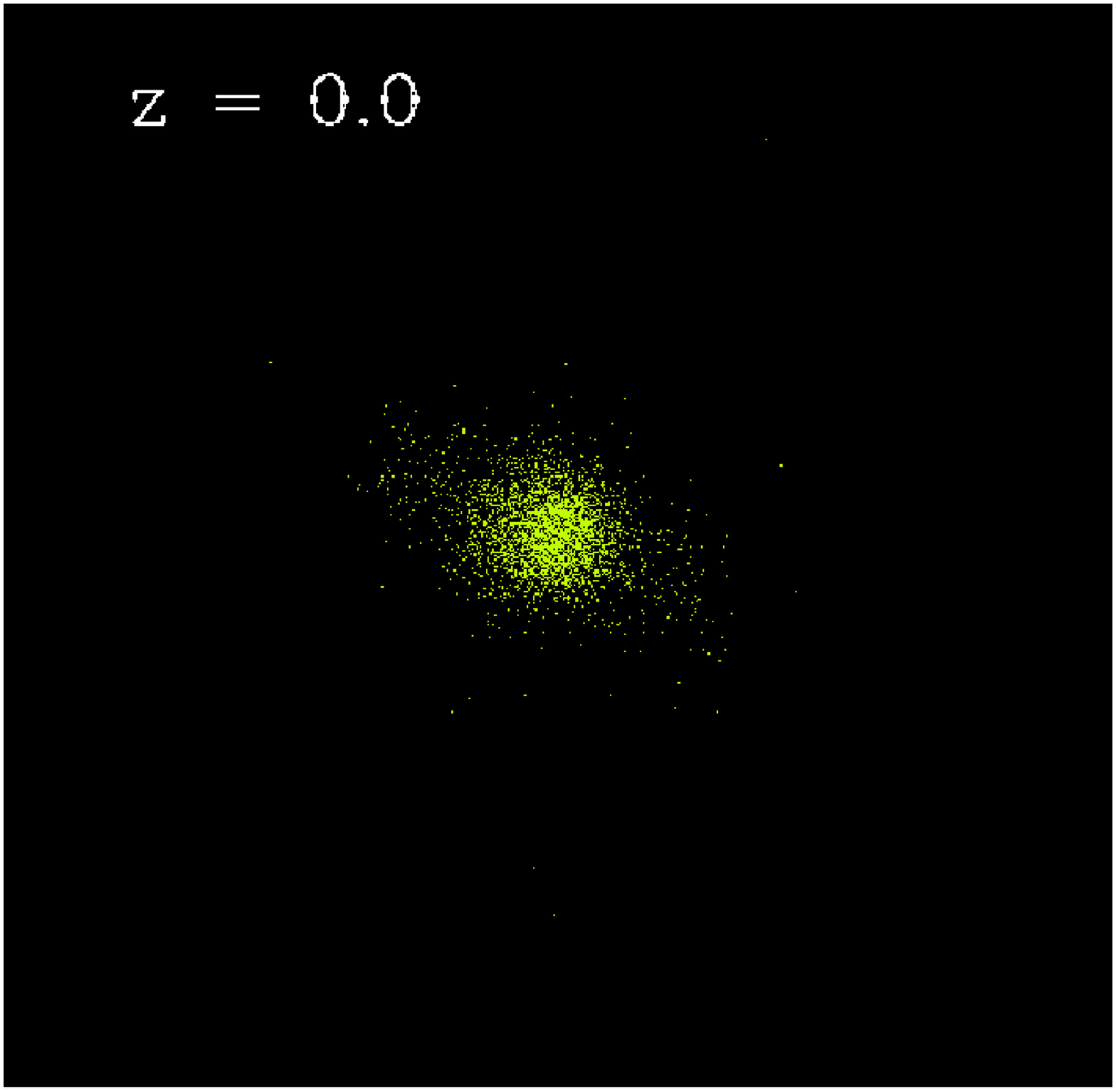}
\includegraphics[angle=0,scale=0.14]{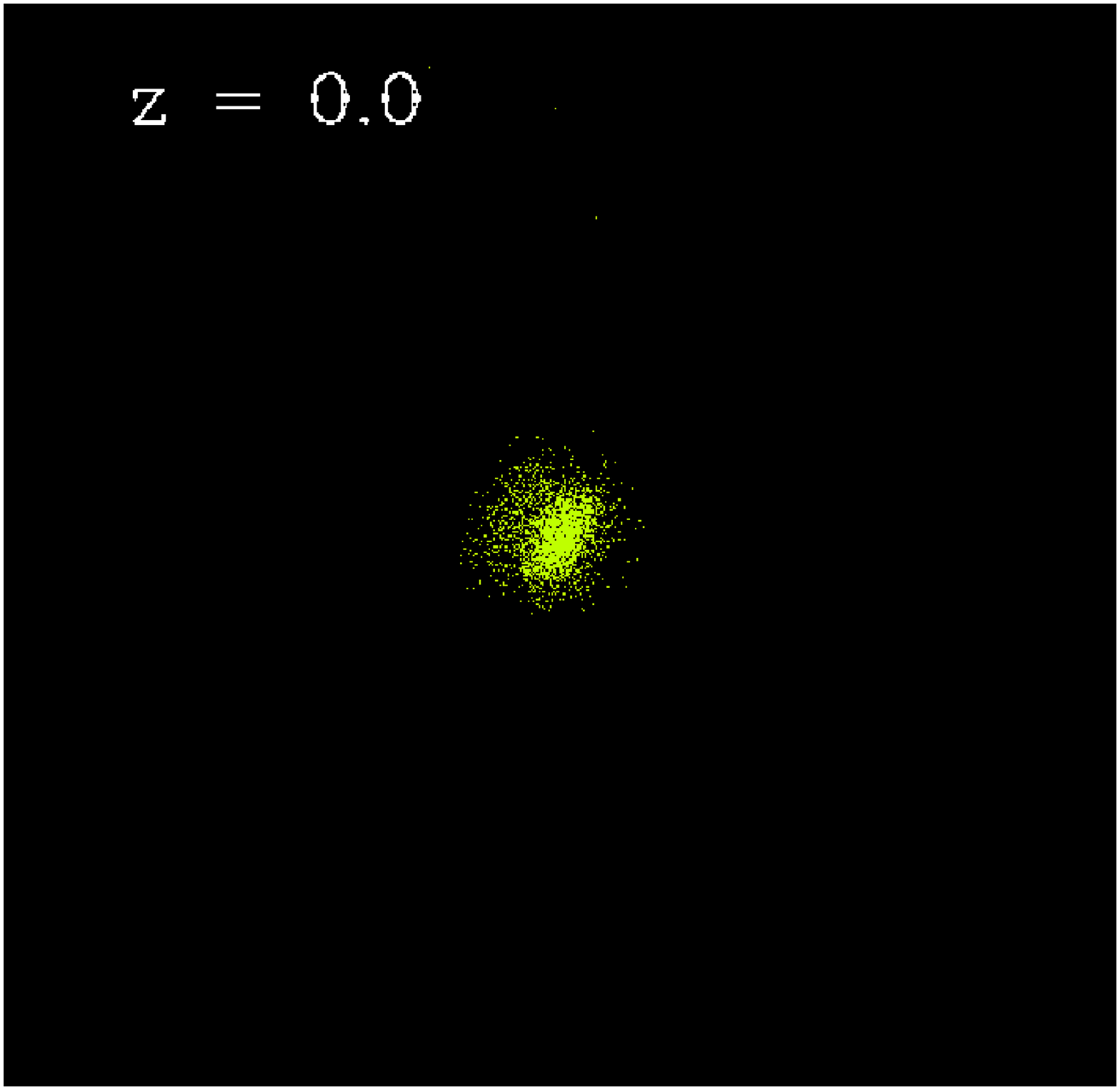}
\includegraphics[angle=0,scale=0.14]{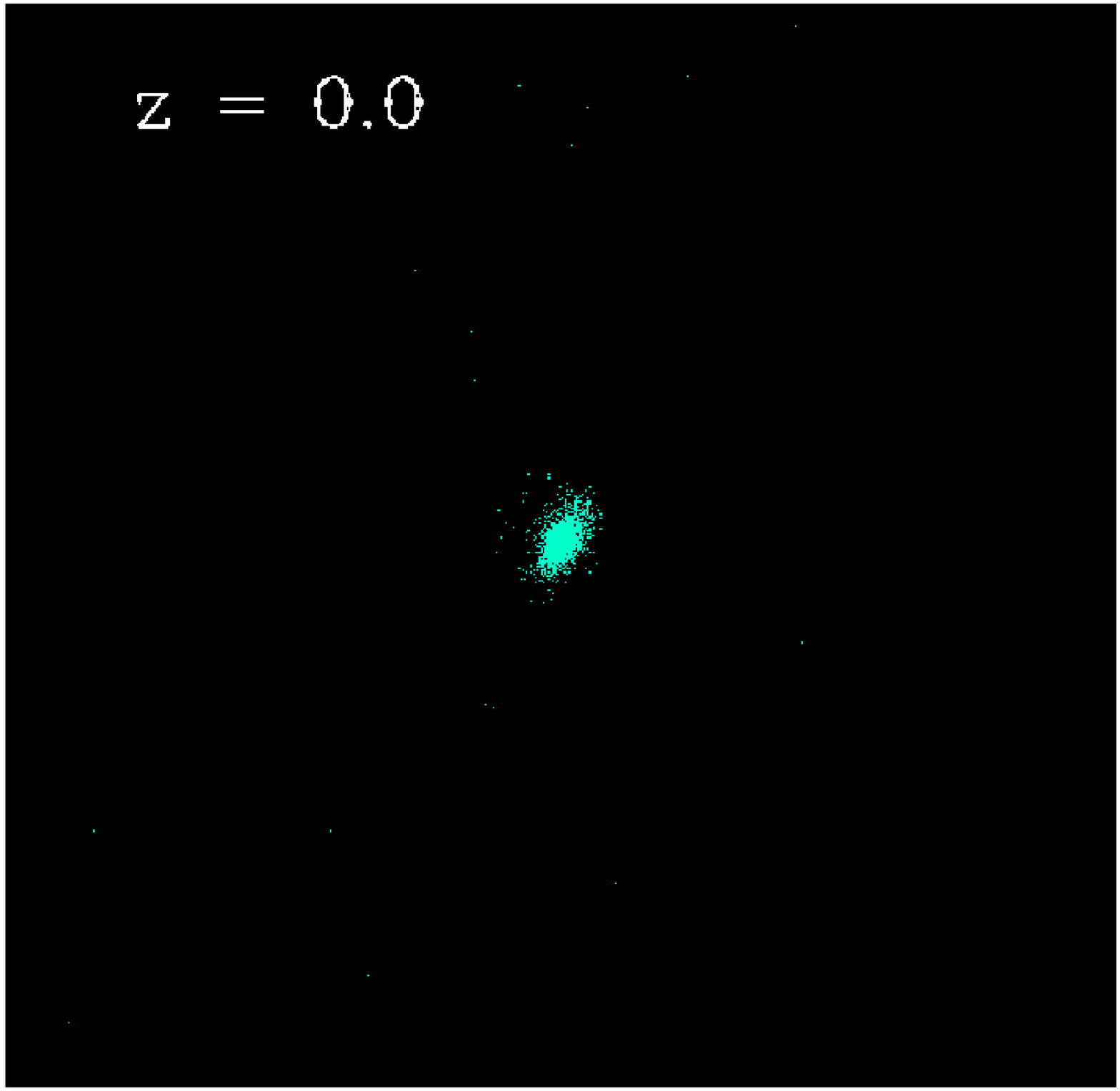}
\includegraphics[angle=0,scale=0.14]{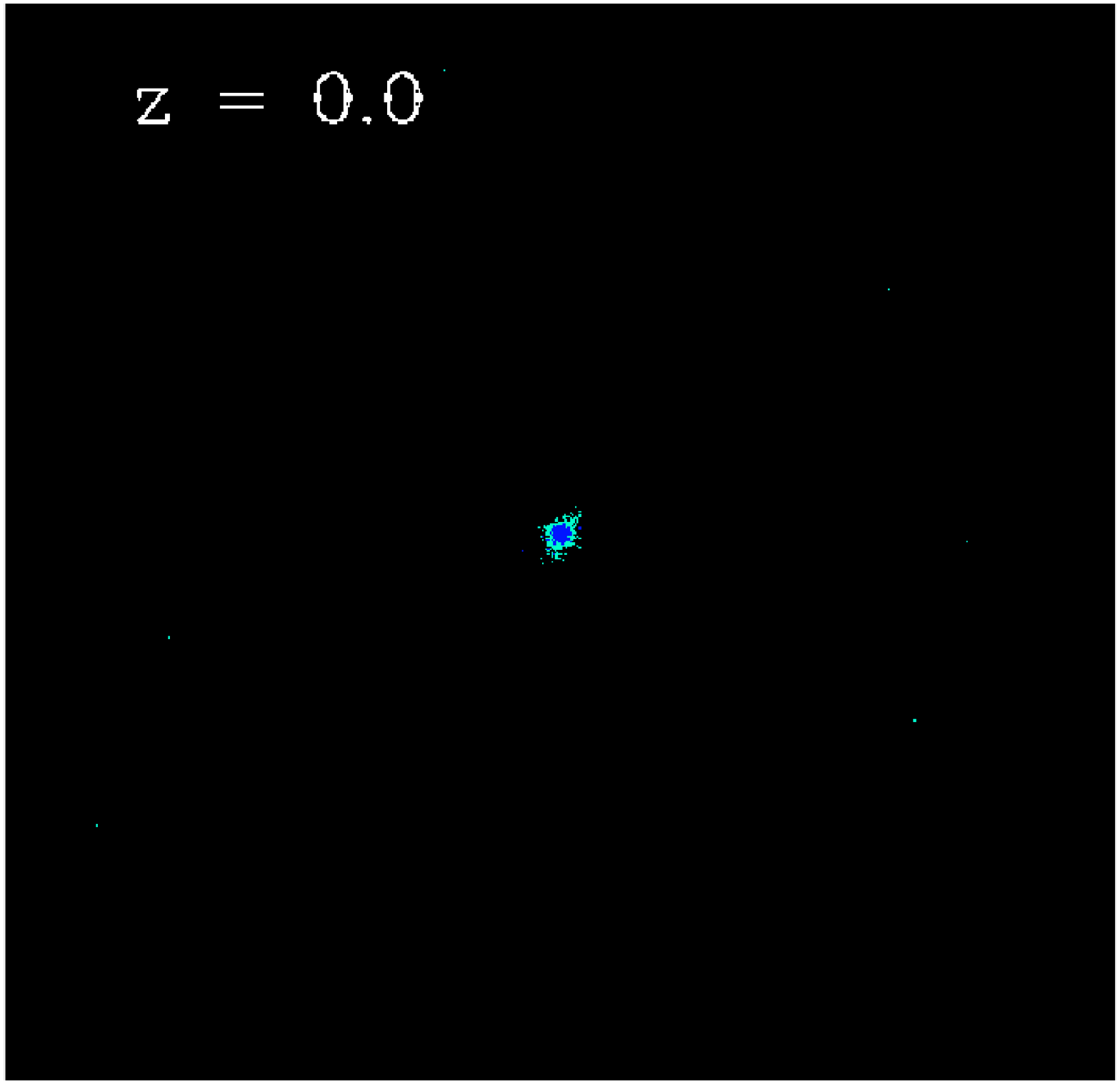}
\end{center}
\caption{Stellar populations of the face-on disk divided into age 
groups at $z=0.3$ (upper) and $z=0$ (lower). From left to right the ages are: 
5--6~Gyr, 3--4~Gyr, 2--3~Gyr, 1--2~Gyr and 0--1~Gyr. The right frames also include
the subgroup of 0--0.03~Gyr old stars. Note the appearance of a {\it nuclear
bar} oriented at $90^\circ$ with respect to the main bar at $z=0$ and visible
for stars younger than $\sim 3$~Gyr (lower frames). Frames: 17~kpc on the side.
}
\end{figure*} 

\begin{figure*}
\begin{center}
\includegraphics[angle=0,scale=0.14]{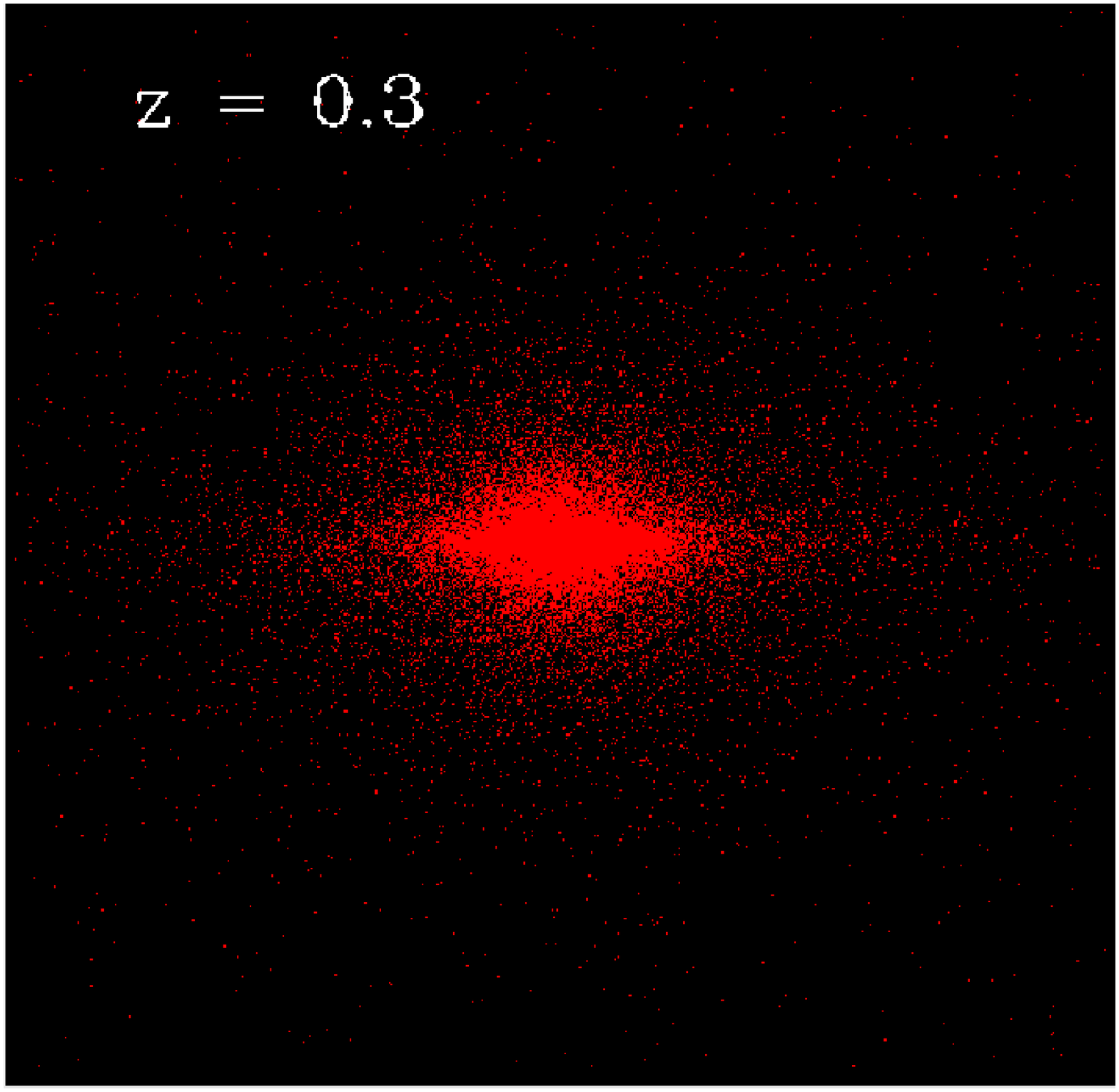}
\includegraphics[angle=0,scale=0.14]{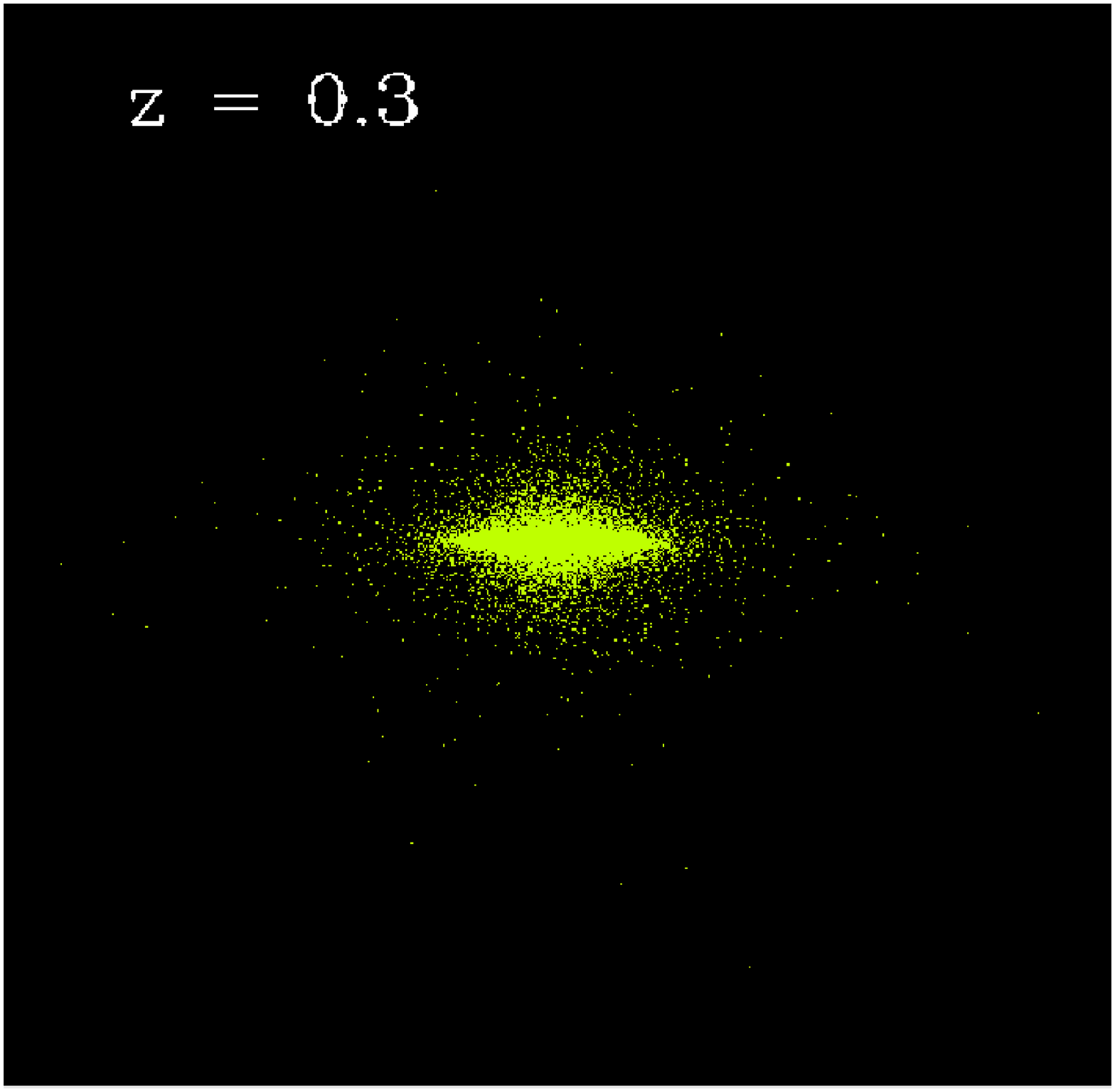}
\includegraphics[angle=0,scale=0.14]{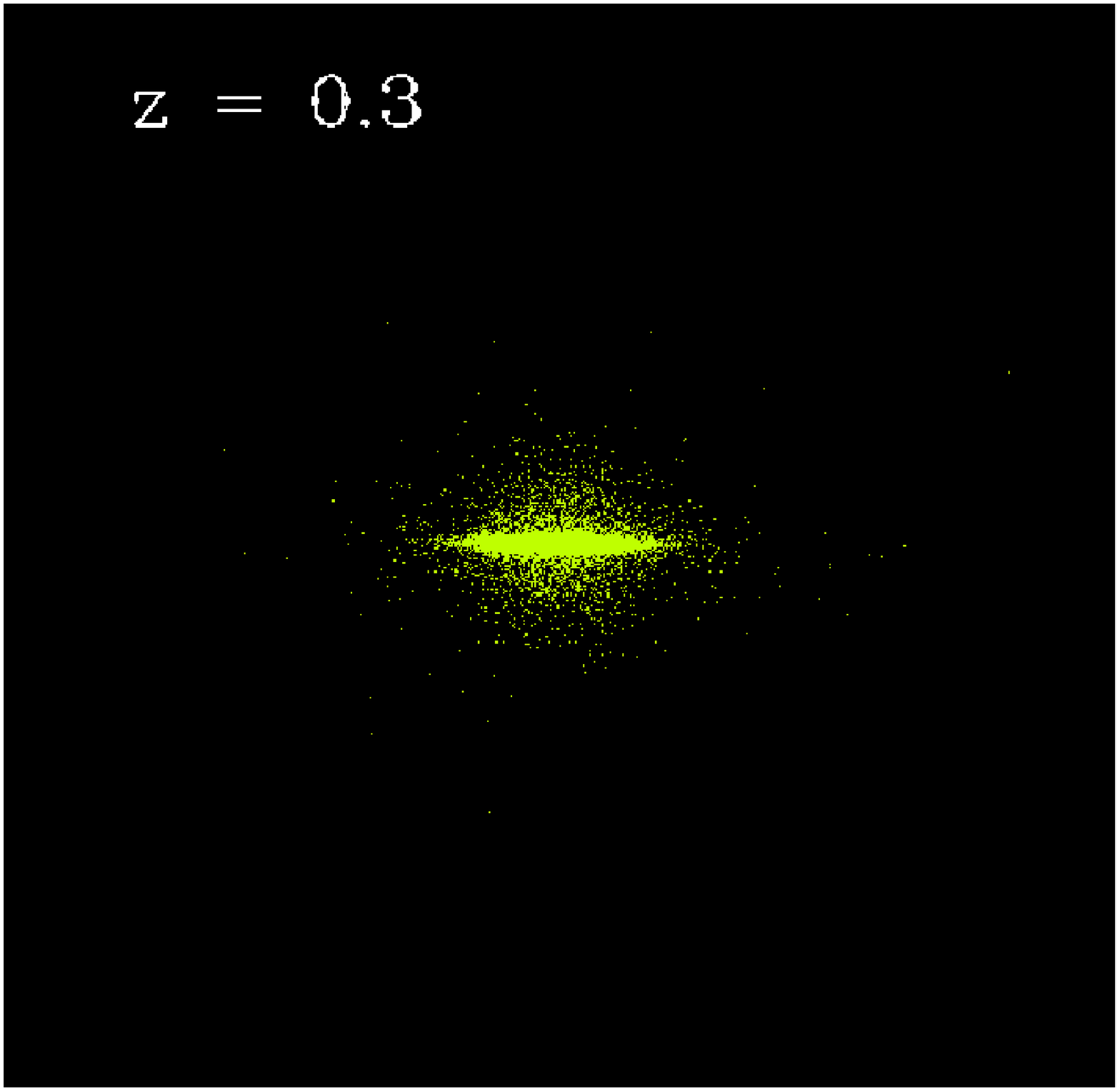}
\includegraphics[angle=0,scale=0.14]{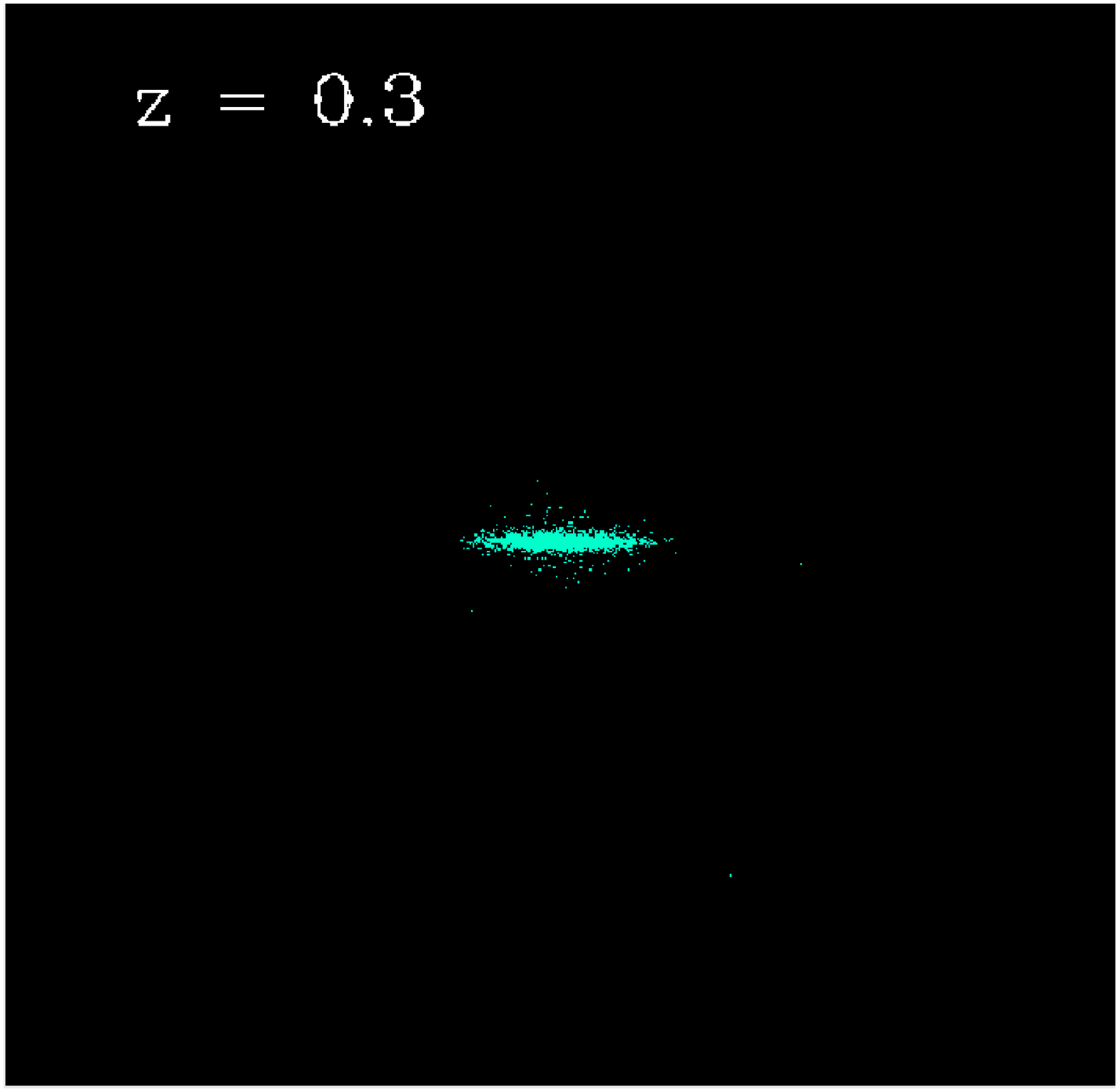}
\includegraphics[angle=0,scale=0.14]{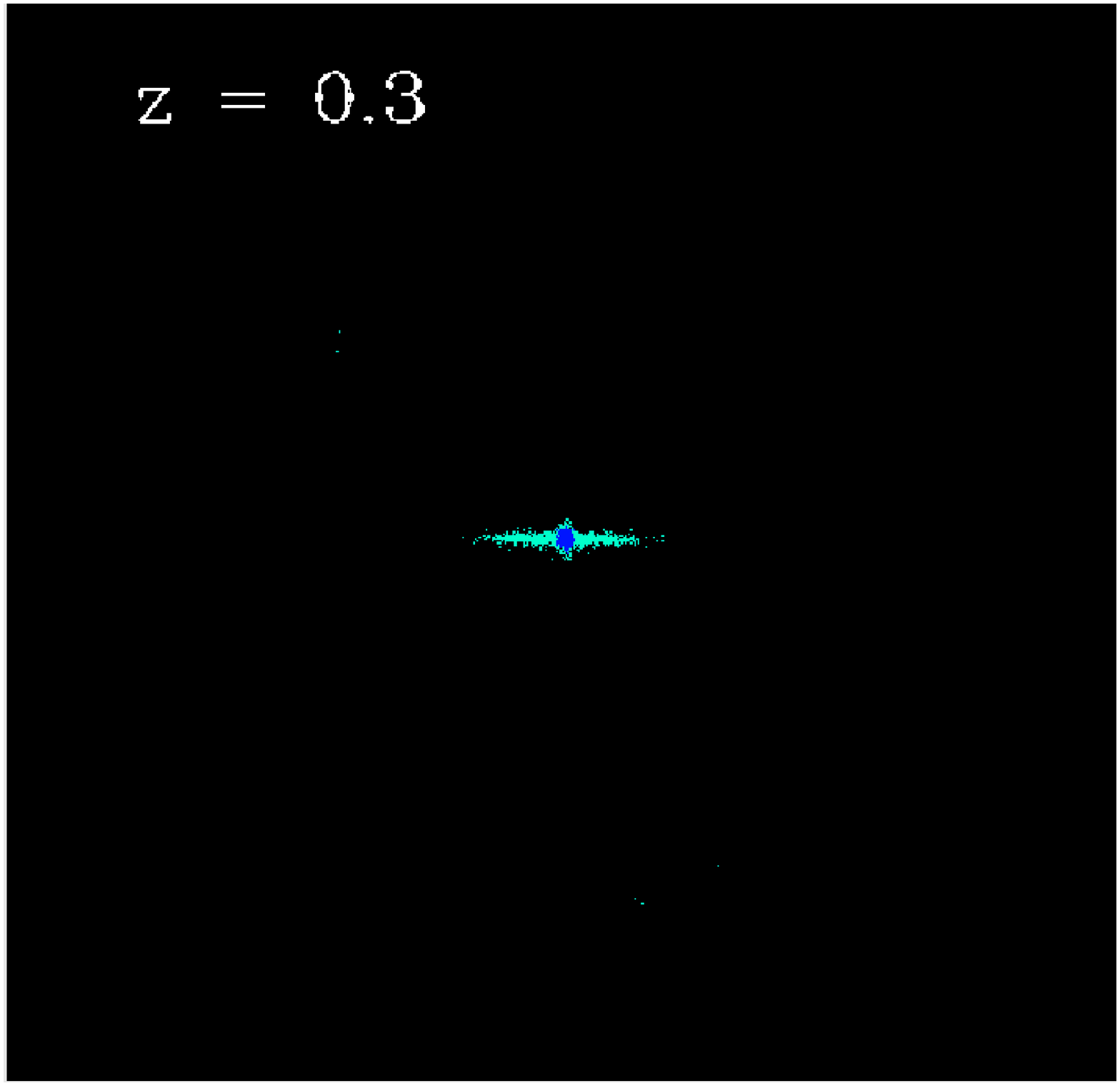}
\includegraphics[angle=0,scale=0.14]{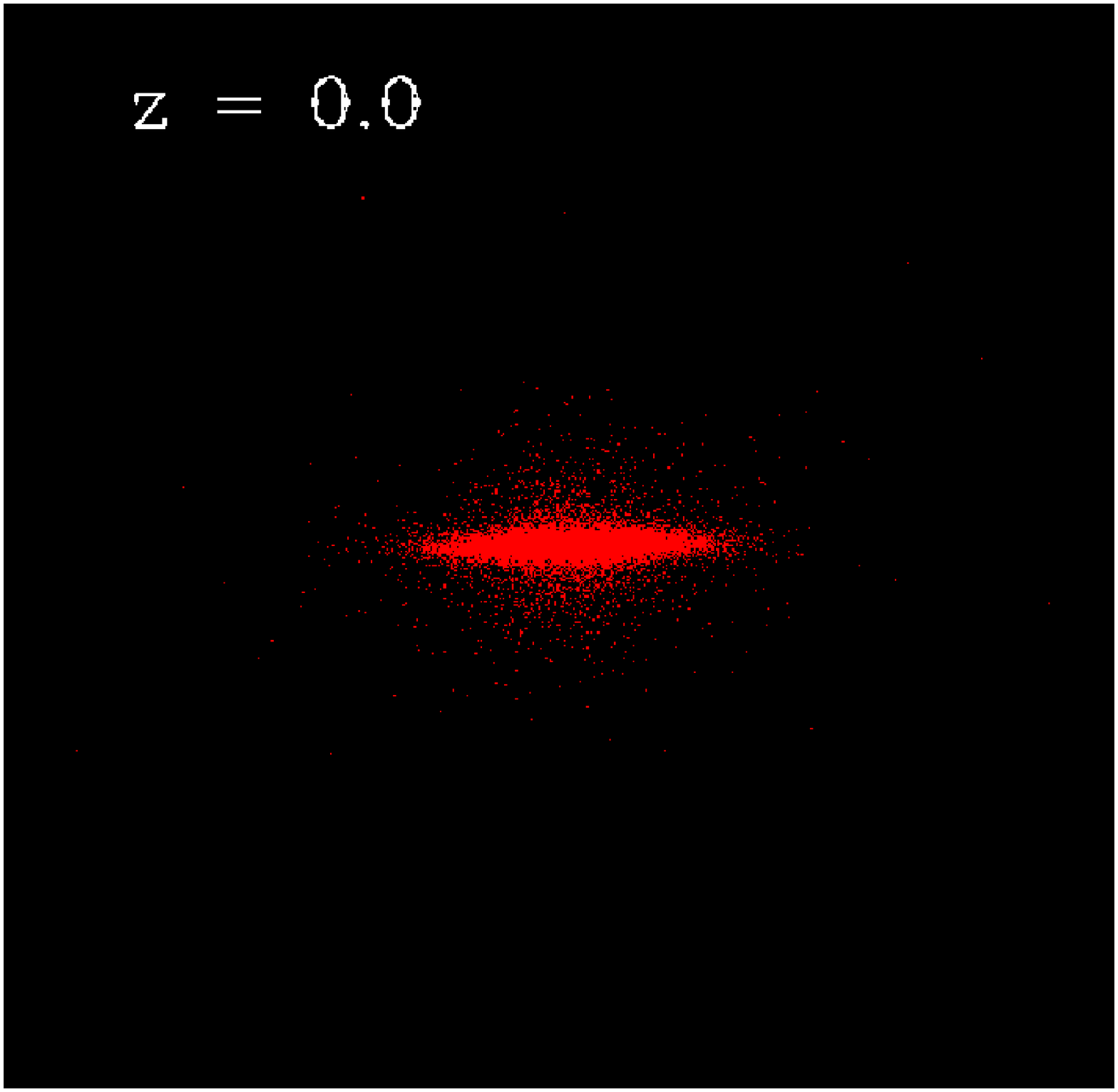}
\includegraphics[angle=0,scale=0.14]{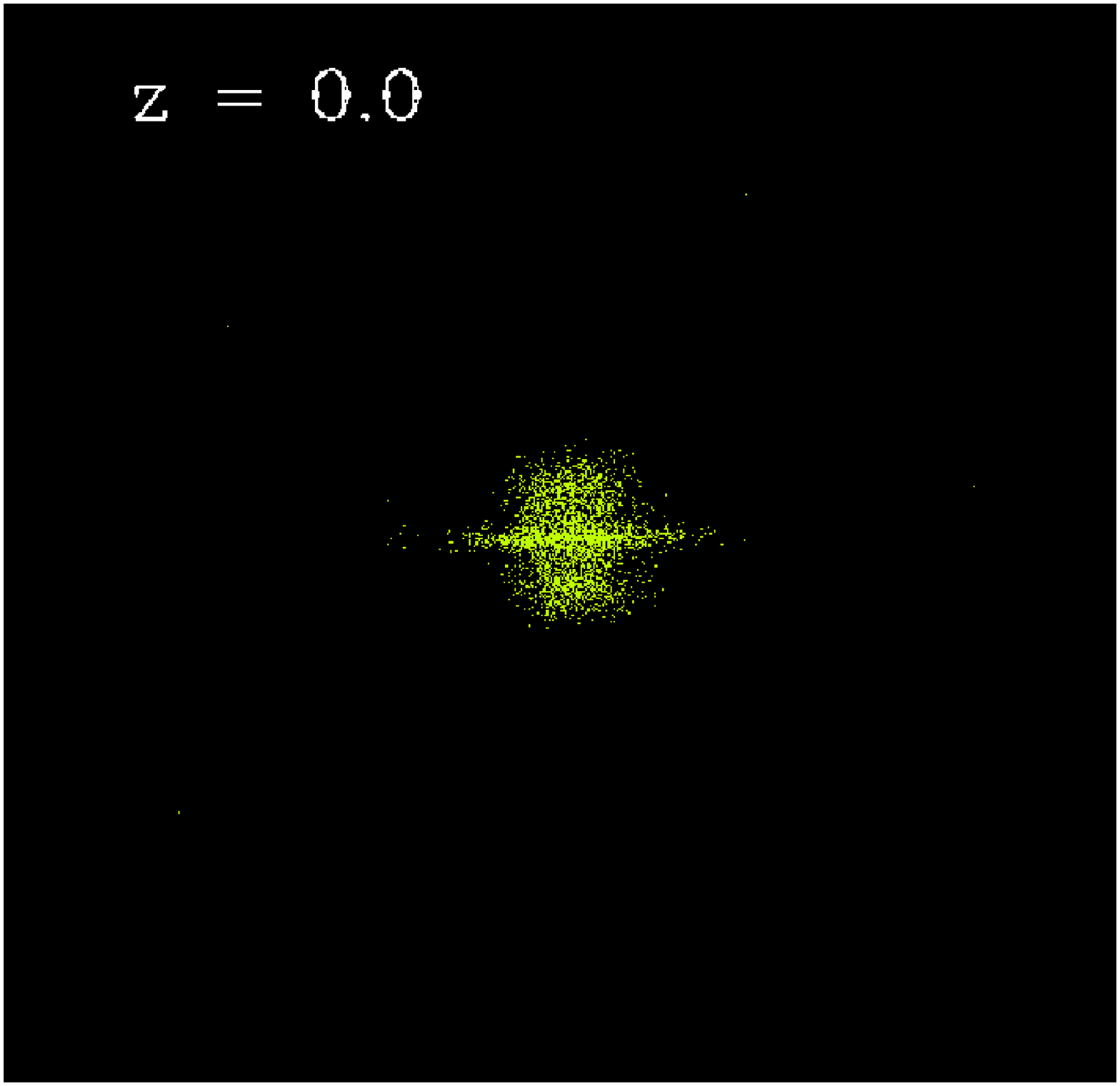}
\includegraphics[angle=0,scale=0.14]{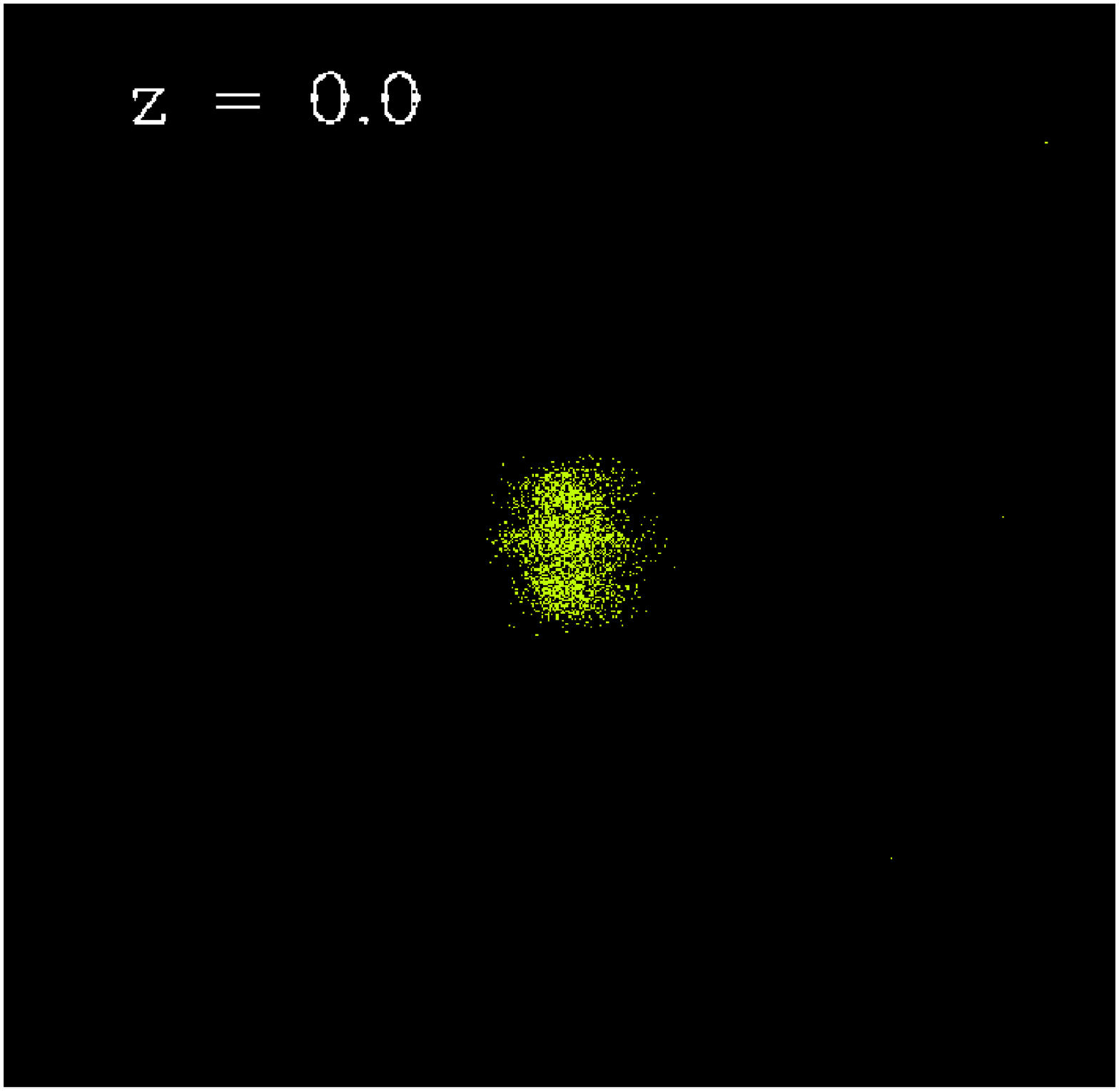}
\includegraphics[angle=0,scale=0.14]{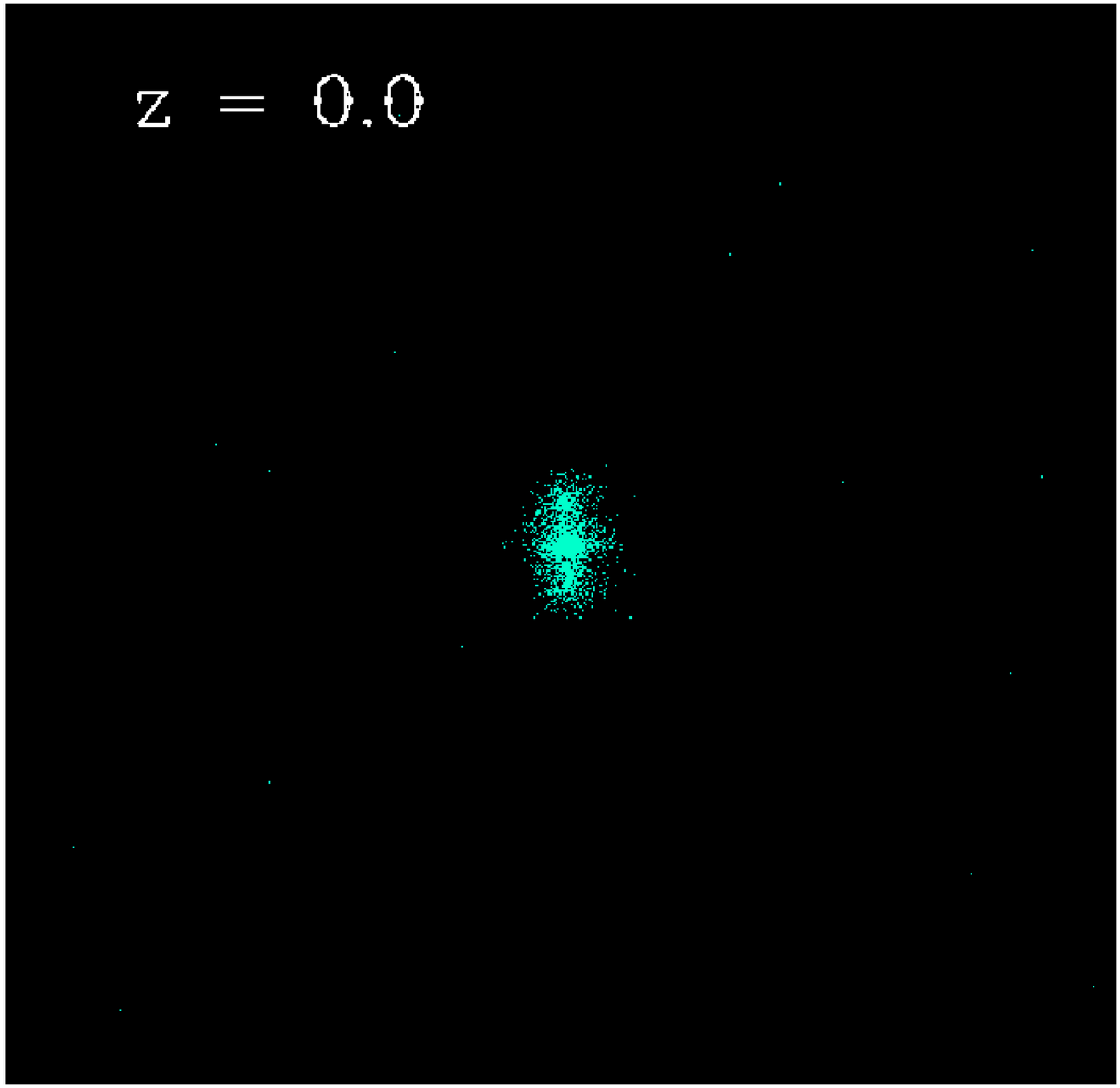}
\includegraphics[angle=0,scale=0.14]{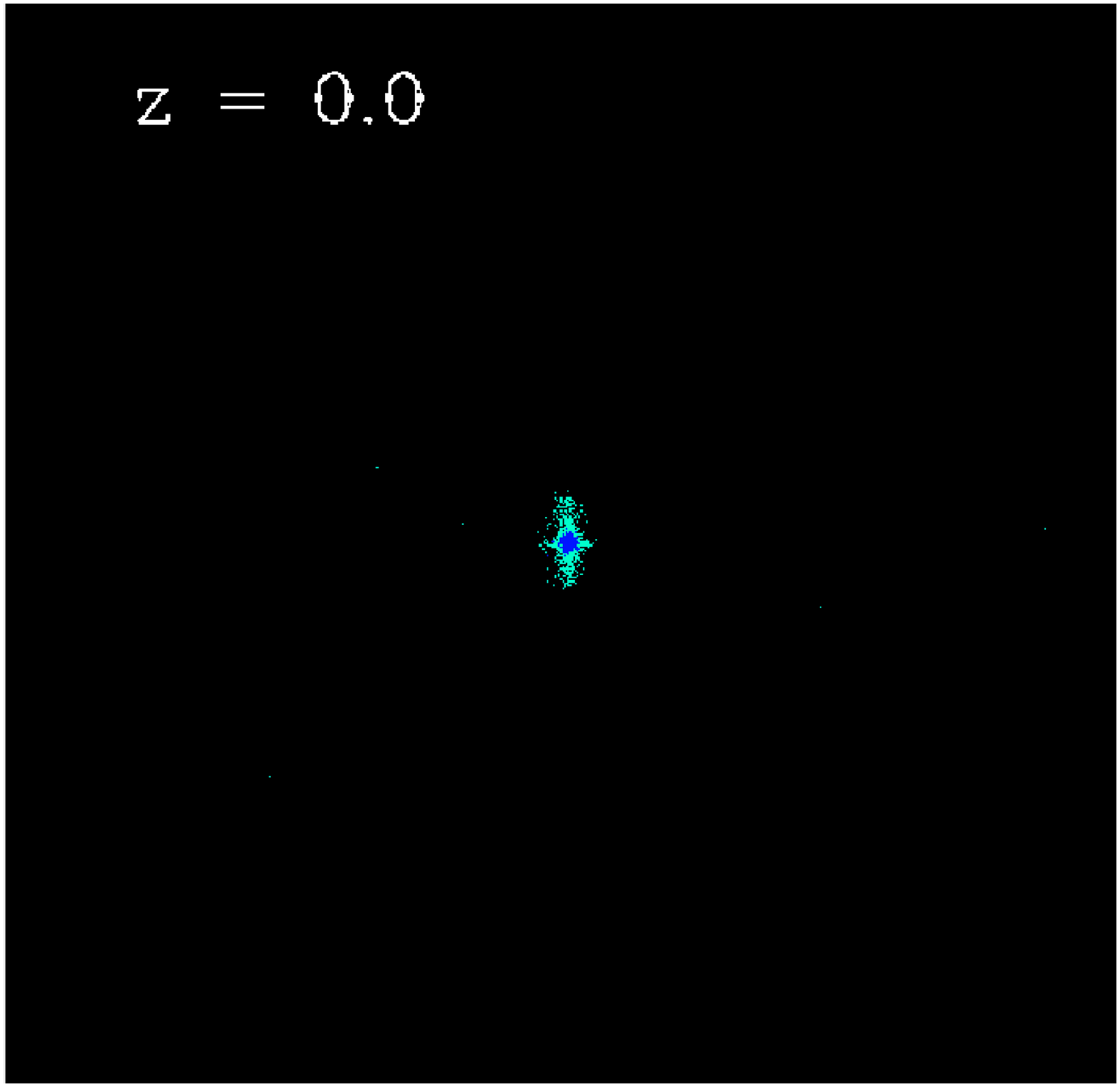}
\end{center}
\caption{Same as Fig.~3 but for an edge-on disk. 
}
\end{figure*} 

For the newly triggered bar, $\Omega_{\rm b}(z)$ largely anti-correlates with 
$A_2(z)$. Amazingly, it increases over $\sim 5$~Gyr until the peak at $z\sim 0.8$,
when $\Omega_{\rm b}$ starts a sharp decay. $\Omega_{\rm b}(z)$ experiences a break at 
$z\sim 0.35$ and becomes shallower, while the bar weakens abruptly (minor merger). 
This evolution is analyzed in \S3.
 
To view the evolution of stellar populations in the bar and the surrounding disk, 
we have separated the stars into age groups: 
5--6~Gyrs, 3--4~Gyrs, 2--3~Gyrs, 1--2~Gyrs, 0--1~Gyrs and 0--0.03~Gyrs old. These 
are shown via frames at $z=0.3$ and $z=0$ for face-on (Fig.~3) and edge-on (Fig.~4)
disks. 

Division into populations reveals certain aspects of the internal dynamics in the 
disk. The older stellar population of 5-6~Gyr dominates the disk and bar 
masses, and the ongoing SF is peaked at the center at $z=0$. 
Figs.~3 and 4 display colors based on stellar ages. The bar has a prominent 
rectangular shape
only when observed face-on in 1--3~Gyr old stars at $z=0.3$. This means that the
bar extends to the ultra-harmonic resonance and its outer part is populated by
stars which are trapped by the characteristic 4:1 orbits.  

A nuclear bar emerges after $z\sim 0.1$
(Fig.~3) for 0--3~Gyr aged stars at $z=0$. One can understand the appearance
of this morphology because the prime bar slows down substantially 
(Fig.~2) and this leads to an inner Lindblad resonance (ILR). A new family of 
periodic orbits forms inside this resonance and 
these orbits are known to be elongated normal to the orbital
family which supports the main bar figure. This new family of orbits is
populated by stars with the lowest dispersion velocities, i.e.,
the youngest stars inside the ILR 
because the trapping ability of the nuclear bar is low. 
 
\section{Discussion: Dynamics and Populations}

We have studied the evolution of galactic bars in fully self-consistent
cosmological simulations of galaxy formation. In a representative model, we 
find that the first generation 
of bars forms early and in response to the asymmetric background DM distribution. This
bar decays quickly and subsequent bars form and are destroyed by interactions
with subhalos. Finally, a long-lived bar is triggered that survives for $\sim 
10$~Gyr. Most interestingly, its pattern speed {\it increases substantially} 
over the first
$\sim 5$~Gyr, despite the angular momentum loss to the outer disk and the 
cuspy DM halo. The bar evolution appears to be closely linked to the disk evolution --- 
it is permeated by interactions with subhalos as we discuss below.

The sequence of bars in the model are repeatedly triggered
by interactions with DM substructure and do not form as a result of a `classical'
(i.e., intrinsic) bar instability. The number of subhalos within the inner prime 
halo is highly variable --- the subhalos cluster
in the filament and are frequently accreted in groups before they merge 
(Romano-Diaz et al. 2008b). This amplifies the damage inflicted by the 
substructure on the disk-bar system because of higher mass and energy influx
into the disk region. We find that the prograde encounters trigger bars,
direct hits weaken existing bars, while retrograde encounters have little effect
--- in line with Gerin et al. (1990) and Berentzen et al. (2004). In spite
of this, addition of the cold gas to the disk, up to $R\sim 15$~kpc, in tandem
with growing stellar surface density by $\sim 10$, allow the bar to speed up
over $\sim 5$~Gyr (Fig.~2). This happens because the buildup of a disk mass
and the increased central mass concentration in the galaxy amplifies the
precession rate of bar orbits.
On the other hand, the weakening of the bar during this time results
from the angular momentum added by the gas to the disk and, therefore, to the
bar. This happens because orbits trapped within the bar become less eccentric,
leading to a weaker bar. 

The gas fraction in the disk exceeds that considered so far in the literature 
by as much as a factor
of 10 (Bournaud et al. 2005; Berentzen et al. 2007). It is an influx of subhalos
into the center which strips the disk of its gas and heats up the stellar
`fluid' injecting it above the plane. Being cut off from the source of angular 
momentum, the bar brakes rapidly.  
  
An important caveat of this slowdown is the bar weakening after $z\sim 0.1$ 
(Fig.~2). Here we do not observe the anticorrelation between $A_2$ and 
$\Omega_{\rm b}$. The appearance of the nuclear bar sheds light on this behavior. 
The orbit trapping by the nuclear bar which grows due to a slowdown deprives the 
main bar of orbital support --- the weakening of the main bar and the 
strengthening of the nuclear bar proceed in tandem.    

Stellar populations provide a further insight into the dynamics of the bar-disk systems.
When viewed in colors of a younger stellar 
population, the disk is thinner and shows less contribution to the spheroidal 
component. Comparison of the left frames in Fig.~4 also shows that the spheroidal 
component is populated by stars older than the average disk population, as it 
disappears in the lower frame. 

\begin{figure}
\begin{center}
\includegraphics[angle=0,scale=0.43]{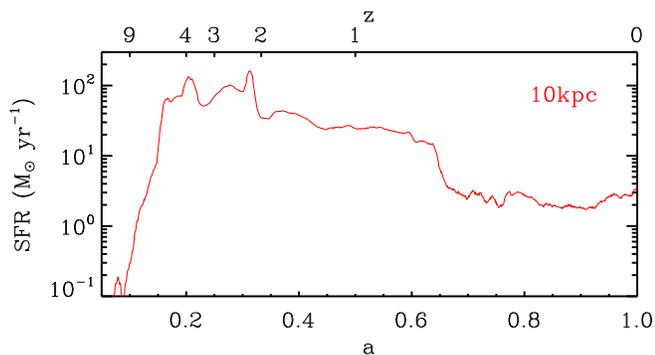}
\end{center}
\caption{SF rates shown within the inner 10~kpc of the disk.
}
\end{figure} 

An interesting feature is present in $z=0$ frames of Fig.~4 with stars younger than
4~Gyr in the center --- the upper/lower surfaces are flat and
parallel to the disk midplane. It is associated with the vertical oscillations
of stars in the region, 
and closely resembles the boxy bulges observed in about 50\% of edge-on disk galaxies 
(e.g., Lutticke et al. 2000). These bulges originate in trapping by a family of 3-D 
orbits belonging to the prime bar (e.g., Combes et al. 1990; Skokos et al. 2002; 
Martinez-Valpuesta et al. 2006).  

The SF continues after the major merger epoch, $z\ltorder 1.5$ (Fig.~5). The 
asymptotic value for the SF rate, SFR$\sim 1~{\rm \msun~yr^{-1}}$, in the 
disk was not fine-tuned but results from self-regulation. Early SF
is supported by the cold gas accretion onto the disk. In the later stage, the SF
is limited to the central kpc only --- this explains the stellar age gradient in the disk
which is positive (Figs.~3, 4). The bar resembles those in the early type galaxies
with a constant surface brightness along the bar major axis, unlike the exponential
bars which show negative age gradients (Perez et al. 2007).

In summary, we have followed the bar evolution in the most realistic environment
attempted so far. We have shown that bars can maintain their fast rotation for
$\sim 5$~Gyr despite the loss of angular momentum to the disk and {\it cuspy} DM halo. 
Finally, bar dynamics appears to be closely linked to the disk stellar populations. 

\acknowledgements
We are grateful to our colleagues, and especially to John Dubinski, Shardha Jogee, 
Seppo Laine, Irina Marinova, Reynier Peletier and Jorge Villa-Vargas for discussions
and comments. This research has been partially supported by NASA and STScI.


\begin{thebibliography}{37}
\expandafter\ifx\csname natexlab\endcsname\relax\def\natexlab#1{#1}\fi


\bibitem[]{}
Berentzen, I., Athanassoula, E., Heller, C.H., Fricke, K.J. 2004,
       \mnras, 347, 220

\bibitem[]{}
Berentzen, I., Shlosman, I., Martinez-Valpuesta, I. \& Heller, C,H. 2007, \apj, 666, 189 

\bibitem[]{}
Bournaud, F., Combes, F. \& Semelin, B. 2005, \mnras, 364, L18

\bibitem[]{}
Byrd, G.G., Valtonen, M.J., Valtaoja, L. \& Sundelius, B. 1986, A\&A, 166, 75 

\bibitem[]{}
Combes, F., Debbasch, F., Friedli, D. \& Pfenniger, D. 1990, A\&A, 233, 82

\bibitem[]{}
Dehnen, W. 2002, J. Comput. Phys., 179, 27

\bibitem[]{}
Diemand, J., Kuhlen, M. \& Madau, P. 2007, \apj, 667, 859

\bibitem[]{}
Dubinski, J., Gauthier, J.-R., Widrow, L. \& Nickerson, S.
     2008, Formation \& Evolution of Disk Galaxies, ASP Conf. Ser., J.G. Funes \& 
     E.M. Corsini (eds.), astroph/0802.3997

\bibitem[]{}
Erwin, P. \& Sparke, L.S. 2002, \aj, 124, 65
     
\bibitem[]{}
Gauthier, G.-R., Dubinski, J. \& Widrow, L. 2006, \apj, 653, 1180

\bibitem[]{}
Gerin, M., Combes, F. \& Athanassoula, E. 1990, A\&A, 230, 37 

\bibitem[]{}
Governato, F., et al. 2004, \apj, 607, 688
  
\bibitem[]{}
Governato, F., et al. 2007, \mnras, 374, 1479
 
\bibitem[]{}
Heller, C.H. \& Shlosman, I. 1994, \apj, 424, 84

\bibitem[]{}
Heller, C.H., Shlosman, I. \& Athanassoula , E. 2007a, \apj, 657, L65

\bibitem[]{}
Heller, C.H., Shlosman, I. \& Athanassoula , E. 2007b, \apj, 671, 226 
 
\bibitem[]{}
Hoffman, Y. \& Ribak, E. 1991, \apj, 380, L5

\bibitem[]{}
Jogee, S., Knapen, J.H., Laine, S., Shlosman, I., Scoville, N.Z. \& Englmaier, P.
   2002a, \apj, 570, L55 

\bibitem[]{}
Jogee, S., Shlosman, I., Laine, S., Englmaier, P., Knapen, J.H., Scoville, N.Z. \&
    Wilson, C.D. 2002b, \apj, 575, 156

\bibitem[]{}
Knapen, J.H., Beckman, J.E., Shlosman, I., Peletier, R.F., Heller, C.H. \& 
    de Jong, R.S. 1995a, \apj, 443, L73

\bibitem[]{}
Knapen, J.H., Beckman, J.E., Heller, C.H., Shlosman, I. \& de Jong, R.S. 1995b, 
     \apj, 454, 623

\bibitem[]{}
Laine, S., Shlosman,I., Knapen, J.H. \& Peletier, R.F. 2002, \apj, 567, 97
     
\bibitem[]{}
L\"utticke, R., Dettmar, R.-J. \& Pohlen, M. 2000, A\&AS, 145, 405    
     
\bibitem[]{}
Martin, P. \& Friedli, D. 1997, A\&A, 326, 449

\bibitem[]{}
Martinez-Valpuesta, I., Shlosman, I. \& Heller, C.H. 2006, \apj, 637, 214  

\bibitem[]{}
Noguchi, M. 1987, \mnras, 228, 635

\bibitem[]{}
Perez, I., Sanchez-Blazquez, P. \& Zurita, A. 2007, A\&A, 465, L9

\bibitem[]{}
Romano-Diaz, E., Hoffman, Y., Heller, C.H., Faltenbacher, A., Jones, D. \&
    Shlosman, I. 2006, \apj, 637, L93

\bibitem[]{}
Romano-Diaz, E., Hoffman, Y., Heller, C.H., Faltenbacher, A., Jones, D. \&
    Shlosman, I. 2007, \apj, 657, 56
  
\bibitem[]{}
Romano-Diaz, E., Shlosman, I., Heller, C.H. \& Hoffman, Y. 2008a, \apj, submitted      

\bibitem[]{}
Romano-Diaz, E., Shlosman, I., Hoffman, Y. \& Heller, C.H. 2008b, \apj Lett., in press,
    October 1 issue, arXiv:0808.0195   

\bibitem[]{}       
Skokos, Ch., Patsis, P.A. \& Athanassoula, E. 2002, \mnras, 333, 847       
       
\bibitem[]{}
Sommer-Larsen, J., G\"otz, M., Portinari, L. 2003, \apj, 596, 47

\bibitem[]{}
White, S.D.M. \& Rees, M.J. 1978, \mnras, 183, 341

\bibitem[]{}Zurita, A. \& Perez, I. 2008, A\&A, 485, 5

\end{thebibliography}


\end{document}